\documentclass{emulateapj}
\usepackage[varg]{txfonts}
\usepackage{graphicx}
\usepackage{float}
\usepackage{longtable}

\shorttitle{Catastrophic outlier prediction in photometric redshift estimation}
\shortauthors{Jones \& Singal}
\slugcomment{{\it Pub. Astron. Soc. Pacific}}

\begin{document}
\title{Tests of catastrophic outlier prediction in empirical photometric redshift estimation\\with redshift probability distributions}

\author{E. Jones\altaffilmark{1} \& J. Singal}
\address{Physics Department, University of Richmond\\138 UR Drive, University of Richmond, VA 23173}
\altaffiltext{1}{Also Department of Physics and Astronomy, University of California, Los Angeles, 430 Portola Plaza, Los Angeles, CA 90095}
\email{evan.jones@astro.ucla.edu}
\begin{abstract}{}{We present results of using individual galaxies' redshift probability information derived from a photometric redshift (photo-z) algorithm, SPIDERz, to identify potential catastrophic outliers in photometric redshift determinations. By using two test data sets comprised of COSMOS multi-band photometry spanning a wide redshift range ($0< z < 4$) matched with reliable spectroscopic or other redshift determinations we explore the efficacy of a novel method to flag potential catastrophic outliers (those galaxies where ${{\vert z_{phot}-z_{spec} \vert}} > 1.0$) in an analysis which relies on accurate photometric redshifts. SPIDERz is a custom support vector machine classification algorithm for photo-z analysis that naturally outputs a distribution of redshift probability information for each galaxy in addition to a discrete most probable photo-z value. By applying an analytic technique with flagging criteria to identify the presence of probability distribution features characteristic of catastrophic outlier photo-z estimates, such as multiple redshift probability peaks separated by substantial redshift distances, we can flag potential catastrophic outliers in photo-z determinations. We find that our proposed method can correctly flag large fractions ($>$50\%)  of the catastrophic outlier galaxies, while only flagging a small fraction ($<$5\%) of the total non-outlier galaxies, depending on parameter choices.  The fraction of non-outlier galaxies flagged varies significantly with redshift and magnitude, however.  We examine the performance of this strategy in photo-z determinations using a range of flagging parameter values. These results could potentially be useful for utilization of photometric redshifts in future large-scale surveys where catastrophic outliers are particularly detrimental to the science goals. } {}
\end{abstract}

\maketitle

\section{Introduction}\label{intro}
Accurate photometric redshift estimates (photo-zs) with well constrained and understood error properties are critical for the current and coming era of large multi-band extragalactic surveys \citep[e.g.][]{Huterer06,Hearin10,BH10}, such as the Large Synoptic Survey Telescope (LSST)\footnote{http://www.lsst.org}, Euclid\footnote{http://sci.esa.int/euclid}, Wide Field Infrared Survey Telescope (WFIRST)\footnote{http://wfirst.gsfc.nasa.gov}, Hyper-Suprime Cam (HSC)\footnote{http://www.naoj.org/Projects/HSC}, and Kilo-Degree Survey (KiDS)\footnote{http://kids.strw.leidenuniv.nl} for which precise redshift estimates will be needed for millions or billions of galaxies extending to high redshifts. In particular, photometric redshift accuracy is the primary source of systematic error in weak-lensing surveys \citep[e.g.][]{BH10}. 

Limiting the occurrence of catastrophic outlier photo-z estimates --- those galaxies whose estimated redshift differs substantially from their actual redshift --- is a top priority for controlling photo-z errors. In addressing this challenge we present a study directed toward a novel method to flag potential catastrophic outlier photo-z predictions through the utilization of individual galaxy redshift probability information. We utilize SPIDERz ---SuPport vector classification for IDEntifying Redshifts \citep{JS17}, a custom implementation of a support vector machine classification model for photometric redshift analysis, which naturally outputs an effective redshift probability distribution for each galaxy\footnote{available from http://spiderz.sourceforge.net with usage documentation provided there.}. SPIDERz's natural output of an effective redshift probability distribution for each galaxy is not necessarily typical for empirical photo-z estimation methods (which make a predictive model based on a training set with known redshifts), but some other empirical methods which can output probability information are ArborZ \citep{Gerdes10}, TPZ \citep{KB13}, SkyNet \citep{Sky}, and ANNz2 \citep{ANNz2}. The techniques discussed in this work should theoretically be relevant to any photo-z estimation method which provides the requisite redshift probability distribution information for individual galaxies.

The performance of candidate photo-z methods should ideally be demonstrated on test data that is representative of the data anticipated by future large-scale surveys. In particular, some data sets, such as much of the LSST catalog, will have photometric data for optical bands only, while others, such as Euclid, will have or overlap with infrared bands. Additionally, some important data sets will span a large redshift range with many high redshift objects. In this work we perform analyses on real data approximating these conditions.  Furthermore, in order to approximate the photometric redshift evaluation conditions of future large-scale surveys, we adopt training set sizes that are much smaller than evaluation set sizes.

\begin{figure}[!htb]
\includegraphics[width=.47\textwidth]{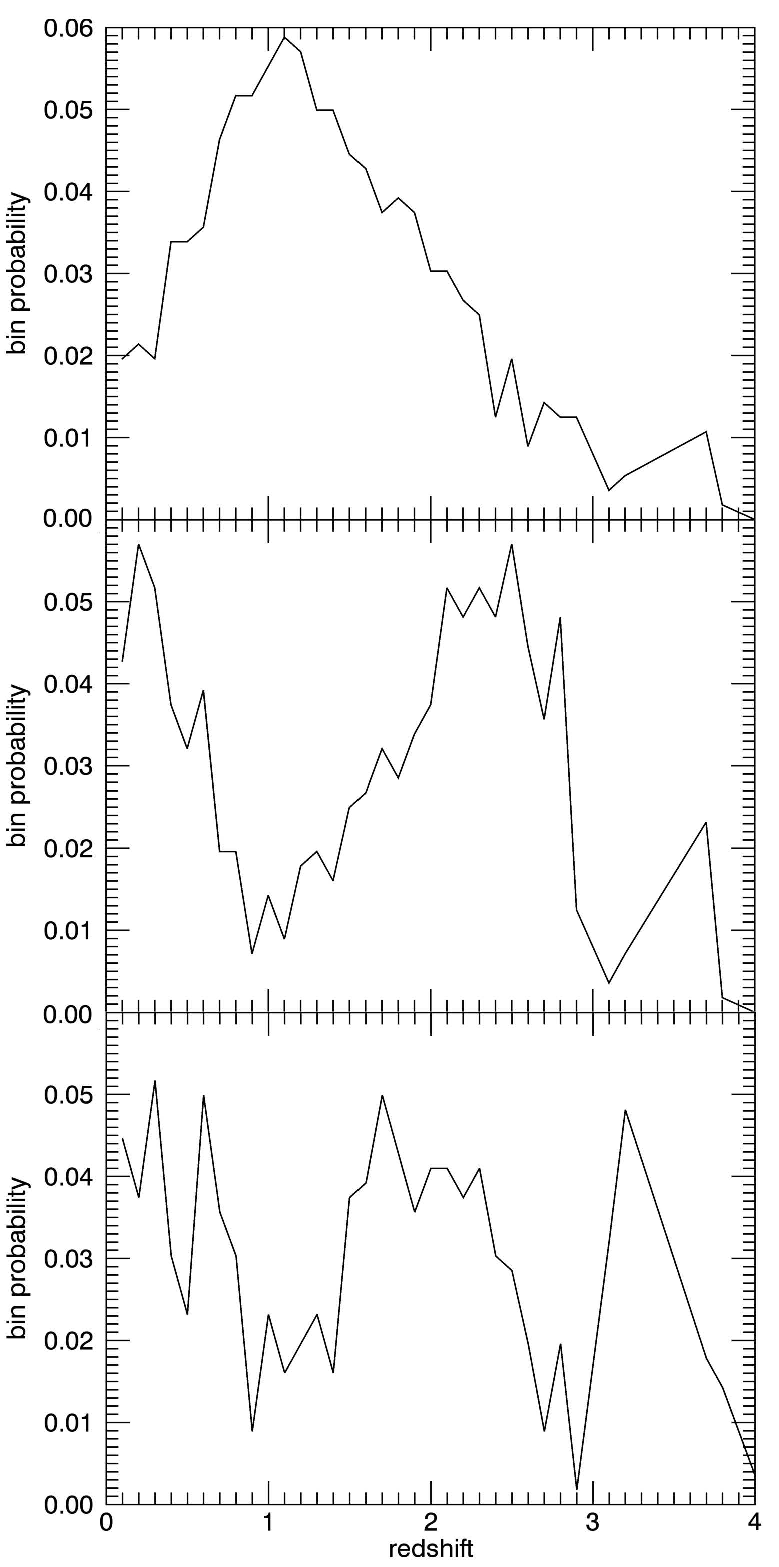}
\caption{Examples of EPDFs as determined by SPIDERz for particular individual galaxies in the COSMOSx3D-HST data set described in \S \ref{data}. The top panel shows an EPDF with a singular uniform probability peak, which is typical of galaxies with accurate redshift estimates. The middle panel shows a classic doubly peaked EPDF where the spectroscopic redshift is near the slightly lower peak, which is often the case for catastrophic outlier redshift estimates. The bottom panel shows an EPDF without a clear probability peak, which also can be the case for catastrophic outlier redshift estimates.}
\label{epdf1}
\end{figure}

Photo-z methods have been traditionally divided into two categories: template-fitting and empirical methods. Template-fitting methods rely on fitting galaxy photometry to template spectra evolved with redshift, typically derived using $\chi ^2$ minimization, e.g. Le Phare \citep{Arnouts99, Ilbert06}, BPZ \citep{Benitez00}, HyperZ \citep{Bolzonella00}, zebra \citep{Feldmann06}, EAZY \citep{Brammer08}, gazelle \citep{KF09}, and DELIGHT \citep{LH17}. Template-fitting methods depend critically on the extent to which galaxy spectral energy distributions (SEDs) library templates adequately represent properties of observed SEDs corresponding to target galaxy populations for which one wants to estimate the redshifts; the selection of ill-fitted SED templates provides the greatest source of errors in redshift determinations with these models. Some techniques for template fitting have incorporated the use of training sets of objects with known photometry and spectroscopic redshifts to better calibrate representative SED templates \citep{Benitez04,Ilbert06,Ilbert09}.

\begin{figure}[!htb]
\resizebox{\hsize}{!}{\includegraphics{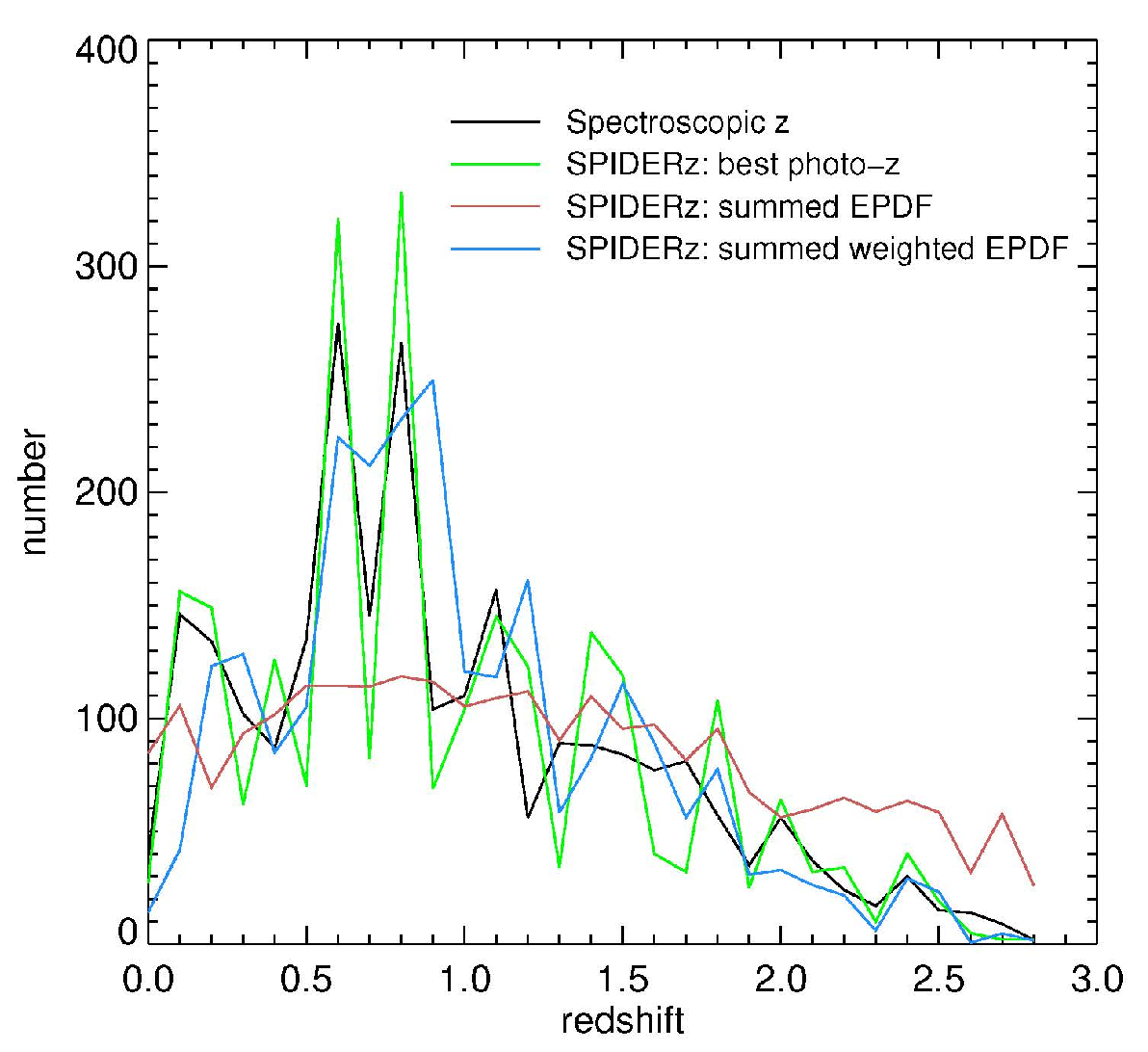}}
\caption{Reconstructed redshift distributions from a determination with SPIDERz using 1200 training galaxies compared to the actual COSMOSx3D-HST evaluation sample of 2323 galaxies. Test data for the determination shown in this figure only were limited to $z < 2.9$ to prevent the occurrence of unoccupied redshift bins at high redshifts. Distributions are shown for the actual spectroscopic redshift, the single best-estimate (highest probability bin) photo-z, the summed EPDF, and the weighted summed EPDF. }
\label{Nz_compare}
\end{figure}

Empirical methods, which rely on training sets with known redshifts to derive a mapping from photometry to redshift, depend critically on the extent to which training galaxy populations adequately represent target galaxy populations in terms of the parameter overlap of photometric inputs and true redshift distributions. Early examples of empirical photo-z methods utilized relatively simple techniques to achieve such a mapping \citep[e.g. polynomial fitting in][]{Connolly95}. More recently, models that produce mappings with greater complexity utilizing machine learning have been examined (e.g. artificial neural networks \citep{Firth03,CL04, Vanzella04,Singal11,Brescia14,Sadeh16}, support vector machines \citep{Wadadekar04,Wang07,JS17}, Gaussian process regression \citep{WS06}, boosted decision trees \citep{Gerdes10}, random forests \citep{KB13,Rau15}, genetic algorithms \citep{Hogan15}, sparse Gaussian framework \citep{Almosallam16}, nearest neighbor search \citep{Ball07, Ball08}, and spectral connectivity analysis \citep{Freeman09}).  A review and comparison of a number of existing photo-z methods can be found in e.g. \citet{Hildebrandt10}, \citet{Abdalla11} and \citet{Sanchez14}.

\begin{figure*}[!htb]
\includegraphics[width=0.5\textwidth]{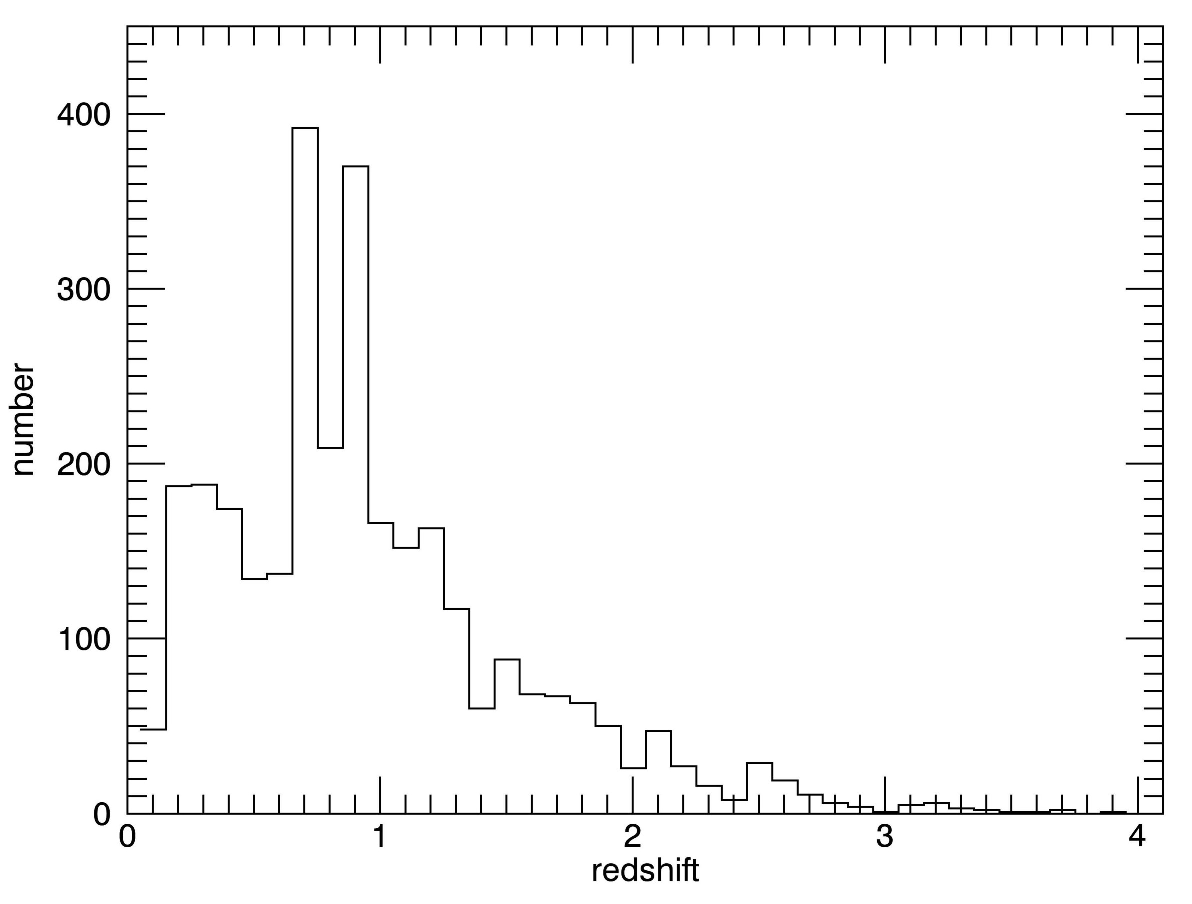}
\includegraphics[width=0.5\textwidth]{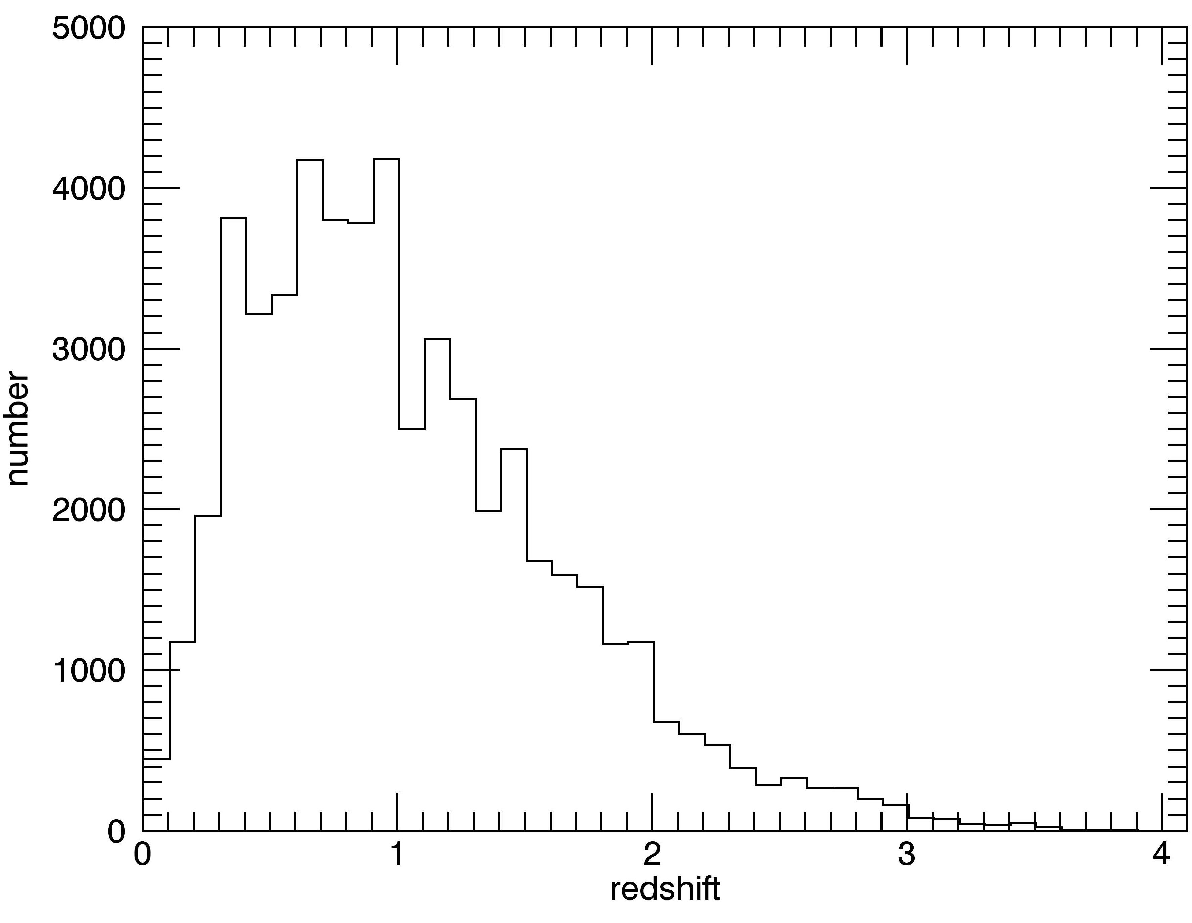}
\caption{$N(z)$ distribution for the 3704 galaxies comprising the COSMOSxHST (left) and 58622 galaxies comprising the COSMOS2015 (right) test data sets used in this analysis.}
\label{p0}
\end{figure*}

Here we follow convention \citep[e.g][]{Hildebrandt10} and define ``outliers'' as those galaxies where
\begin{eqnarray}
Outliers: {{\vert z_{phot}-z_{spec} \vert} \over {1+z_{spec}}} > .15,
\label{erroreq}
\end{eqnarray}
where $z_{phot}$ and $z_{spec}$ are the estimated photo-z and actual (spectroscopically determined) redshift of the object. Although there is not a standard, universal definition of ``catastrophic outliers'' we use a definition that is typical \cite{BH10}:
\begin{eqnarray}
O_{c}: {{\vert z_{phot}-z_{spec} \vert}} > 1.0.
\label{erroreqCO}
\end{eqnarray}
The RMS photo-z error in a realization is given by a standard definition
\begin{eqnarray}
\sigma_{\Delta z/(1+z)} \equiv \sqrt { {{1} \over {n_{gals}}} \Sigma_{gals} \left( {{ z_{phot}-z_{spec} } \over {1+z_{spec}}} \right) ^2 },
\label{RMSeq}
\end{eqnarray}
where $n_{gals}$ is the number of galaxies in the evaluation testing set and $\Sigma_{gals}$ represents a sum over those galaxies.
We also calculate the RMS error without the inclusion of outlier galaxies, referring to this quantity as the ``reduced'' RMS or R-RMS.

In \S \ref{model} we present a summary overview of the SVM model implemented in SPIDERz and discuss the probability information produced for each galaxy. In \S \ref{gen} we present a method for flagging potential catastrophic outlier photo-z estimates made by SPIDERz through the utilization of redshift probability information. In \S \ref{res} we discuss the results of testing SPIDERz on the two test data sets utilizing COSMOS multi-band photometry. We present a discussion in \S \ref{disc}.

\section{SPIDERz and effective probability distributions}\label{model}

A full discussion of the SPIDERz algorithm, mathematical theory, and a suite of tests with various data sets and comparisons with other photo-z determination methods is available in \citet{JS17}. Here, we will provide a brief outline of the machine learning photo-z process for context, but we primarily focus on the utilization of the naturally available probability information for each galaxy produced during photo-z evaluations with SPIDERz. The general technique we propose in this work for utilizing the probability information, however, should theoretically be relevant to any photo-z estimation method which provides the requisite probability information for individual galaxies.

Generally speaking, machine learning photo-z codes perform two main processes: training and evaluation. The output of the training process is a mapping from band magnitudes (and potentially additional information) to redshift. The collection of mappings comprise a predictive model that can be used to make photo-z predictions on evaluation galaxies.

SPIDERz utilizes support vector classification to make photo-z predictions, where bins of redshift are assigned class labels, and photo-z estimation is performed via the solution to a multi-class classification problem. SPIDERz solves the multi-class problem with a ``one against one'' or ``pairwise coupling'' approach that treats the complex multi-class problem as a series of simpler binary class problems consisting of every possible pairing of classes (in this case redshift bins). Thus for a system comprised of $m$ distinct classes ($m$ redshift bins in this case), SPIDERz formulates and solves $m(m-1) \over 2$ separate binary classification problems, choosing the more likely class (redshift bin) in each binary pairing. Each instance of classification in favor of a particular redshift bin can be regarded as a `vote' for that class. The entire collection of $m(m-1) \over 2$ votes forms a distribution (see Figure \ref{epdf1} for examples) that we call an `effective' probability distribution (EPDF) for each galaxy, with the relative probability of each redshift bin proportional to the number of times the corresponding class was chosen as the best binary solution. This EPDF is not continuous, but rather is resolved to the bin-width level. Discrete $z_{phot}$ estimates, if they are desired, can be obtained for each galaxy by simply taking the redshift bin with the highest number of votes.

Examples of actual EPDFs for individual galaxies in the COSMOSx3D-HST data set described in \S \ref{data} are shown in Figure \ref{epdf1}. The top panel shows the presence a uniform singular probability peak characteristic of typical cases where $z_{phot} \approx z_{spec}$. The middle and bottom panels show distributions with multiple peaked probabilities throughout wide redshift distances, which is a feature that is typical of many inaccurate $z_{phot}$ estimates.

We use the terminology ``effective PDF'' because of the way that all bins are used in comparison, thus artificially inflating low probability bins due to the inevitable pairwise comparisons of two low probability bins. However the overall shape of the EPDFs, in regard to higher probability bins which are the only ones relevant in this analysis, approaches that of a true probability distribution.  

To briefly illustrate how the EPDF compares to a true probability distribution function, if one desires to mitigate the effect of low probability bin inflation in the EPDFs for comparisons between the summed EPDF for all galaxies and the known $N(z)$ distribution in testing determinations, one would apply weights to the EPDFs that are proportional to the fractional population of training galaxies in each redshift bin relative to the total training galaxy population. Weights are determined for each redshift bin $\Delta z_i$ by
\begin{eqnarray}
w_i = {{N(\Delta z_i)} \over {N}}
\label{weights}
\end{eqnarray}
\begin{eqnarray}
\displaystyle\sum_{i=1}^{l} w_i = 1,
\label{weightsnorm}
\end{eqnarray}
where $N = \displaystyle\sum_{i}^{l} N(\Delta z_i)$ and $l$ is the number of redshift bins. Weights are applied to the EPDF by

\begin{eqnarray}
P_{w}(\Delta z_i) = l(\displaystyle\sum_{i=1}^{l} P_{Eff}(\Delta z_i)* w_i)),
\label{EPDFweighted}
\end{eqnarray}
where $P_{w}(\Delta z_i)$ is the weighted probability for some redshift bin $\Delta z_i$, and $P_{Eff}(\Delta z_i)$ is the probability given by the unweighted EPDF.  In this way, as shown in Figure \ref{Nz_compare}, we can see that there is meaningful probability information in the EPDFs and that they can be made, in aggregate, to approach a true probability distribution with weighting.  For the present work, however, the degree of fidelity of the EPDFs to true probability distribution functions is not important, as only the highest probabilty bins are relevant, and so no weighting is applied --- the analyses in this work simply use the raw EPDFs as output by SPIDERz.  The reason for this is severalfold:  firstly, we would like to demonstrate the method of this work with the raw output of a machine learning classifier, for the simplest, most general situation.  Further, while it is the case that in the analyses here the training set and the evaluation set have practically the same redshift distribution, that is not necessarily the case for all generic photo-z evaluations going forward, so weighting the individual output galaxy probabilities by the particular redshift distribution of the training set may not be appropriate.  Additionally, in this work we are focusing on the utility of individual galaxy probability functions.  If one were to weight those functions individually by the cumulative redshift distribution of a given single training set, the amount that high probability peaks are scaled up and down would be highly dependent on the particulars of that training set, and would be different for another training set; therefore values investigated quantitatively here would be entirely training set dependent, and certain training sets would result in a weighting where no individual galaxies have high probability peaks at high redshifts.

We note that to produce Figure \ref{Nz_compare}, due to the relatively limited population of galaxies at high redshifts in the COSMOSx3D-HST data set used in this analysis, the presence of unpopulated redshift bins at high $z$ in a training set is often unavoidable.  So in order to present a useful comparison between the summed EPDFs and distribution of discrete most probable estimates $N(z_{phot})$ produced in SPIDERz determinations with the actual redshift distribution $N(z_{spec})$ for this particular data set, we utilized a subset of test data galaxies restricted to $z < 2.9$, ensuring all redshift bins are populated for this particular calculation only.

By default, SPIDERz chooses the most probable (commonly occurring) redshift bin as a single valued photo-z estimate for the galaxy. In this analysis we use this method for discrete photo-z predictions, such as those shown in Figure \ref{f1_removed_red_xs}. In this work we seek a method to identify potential catastrophic outliers in such photo-z predictions.

SPIDERz also allows users flexibility in redshift bin size. We generally find determinations have increased accuracy and precision when smaller bin sizes are used, however the optimal bin size for any determination will be dependent on the size and nature of the training set (decreasing the bin size for determinations lowers existing parameter overlap between training and evaluation sets), and can be approached via trial-and-error or approximated with the bin size introduced as an additional parameter in a grid search (see a detailed discussion in \cite{JS17}).

\section{Strategy for identifying potential catastrophic outliers with EPDFs}\label{gen}

To identify potential catastrophic outlier photo-z estimates we focus on the existence of individual galaxies' EPDFs displaying multiple probability peaks, or somewhat equivalently, a `weak' primary probability peak. There is some ambiguity in what constitutes multiple substantial probability peaks in a galaxy's EPDF. In particular, a secondary peak is more likely to be significant if it is closer in height (probability) to the primary (highest probability) peak, and also if it is located farther away in redshift from the primary peak. Let us denote the ratio of the probability of a secondary peak $i$ to the primary peak in a galaxy's EPDF as
\begin{equation}
p_f = {{p_{i}} \over {p_{max}}},
\end{equation}
where $p_{max}$ is the probability of the primary (highest probability) peak, and let us also denote the redshift distance between that secondary peak and the primary peak as $\Delta{z_{peak}}$. Thus a designated minimum value for $p_f$ ($p_{f,min}$), and a designated minimum value for $\Delta{z_{peak}}$ ($\Delta{z_{peak,min}}$), can serve as filter values above which a multiply peaked EPDF is flagged. If at least one redshift bin in an EPDF distribution satisfies both of the $\Delta{z_{peak}} > \Delta{z_{peak,min}}$ and $p_f > p_{f,min}$ criteria, the galaxy is flagged as a potential catastrophic outlier. The optimal values for $p_{f,min}$ and $\Delta{z_{peak,min}}$ will vary depending on factors such as the redshift range of test data and designated bin size, and the relative importance of flagging more catastrophic outliers versus avoiding spurious flaggings.

The simplest way to deal with flagged galaxies would be to remove them from analyses which rely on photo-zs. This would, of course, remove some fraction of catastrophic outliers and other outliers, along with some fraction of non-outliers. In \S \ref{COSMOSres} we show that the former number can be relatively high and the latter relatively low. In this analysis going forward we consider flagging as somewhat equivalent to removal from consideration, while acknowledging that other strategies, such as de-weighting while not completely eliminating flagged galaxies in analyses, are possible and likely desirable in some circumstances.

\section{Results}\label{res}

In this section, we present the results from our study of using EPDFs to identify probable outlier and catastrophic outlier galaxy estimates as discussed in \S \ref{gen}. We begin with a discussion of the two test data sets used in these photo-z analyses. Next we provide results from photo-z determinations performed with SPIDERz on the test data sets --- both with and without application of the EPDF outlier identification method discussed in \S \ref{gen}. Metrics of performance for this method are provided for a range of values for the identification criteria, assuming here a simple removal of flagged galaxies.

\subsection{Test Data Sets}\label{data}

To obtain a data set of real galaxies with publicly available spectroscopic redshifts containing sources throughout a large redshift range including higher redshifts we use spectroscopic redshifts from the 3D-HST survey performed with the Hubble Space Telescope and reported in \citet{Momcheva16} that overlap with photometry from the COSMOS2015 photometric catalog \citep{Laigle15} which reports photometry for over half a million objects in the COSMOS field \cite{Scoville07}.  For spectroscopic redshifts we use the reported ``best available'' redshift measurement and eliminate those flagged as having their redshift obtained from photometry or as being stars.  This results in a data set of 3704 galaxies, of which 383 (10.3\%) have $z > 2$ and 948 (25.6\%) have $ z > 1.5$. The $N(z)$ distribution for this data set is shown in Figure \ref{p0}. These data span an $i$-band magnitude range from 27.05 to 18.16 with a median of 23.74.

\begin{figure}[!htb]
\includegraphics[width=0.47\textwidth]{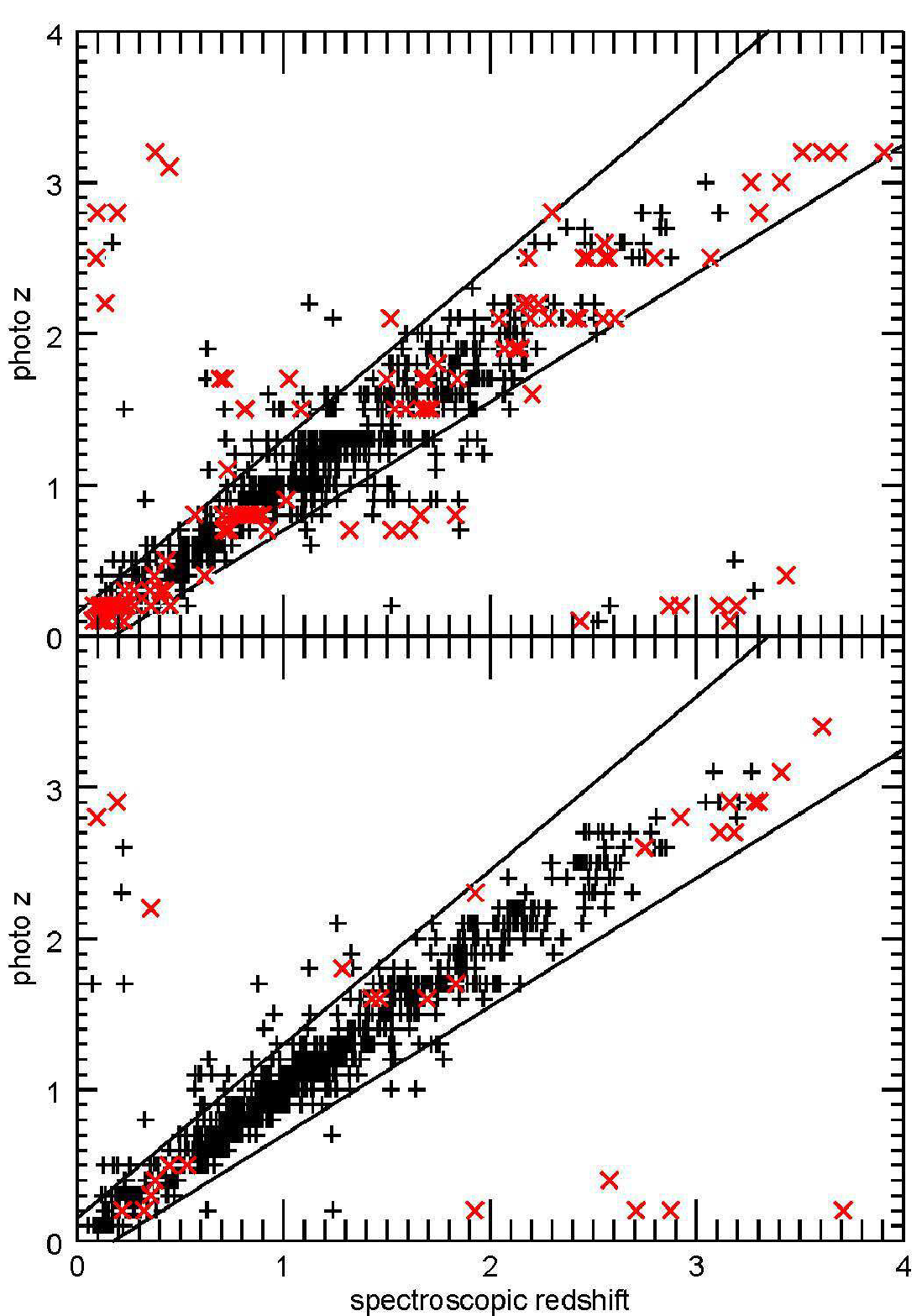}
\caption{The best discrete photo-z estimation (most probable redshift, as discussed in \S \ref{model}) as determined by SPIDERz versus the actual redshift for the COSMOSx3D-HST data set discussed in \S \ref{data} for a realization of the five-band ($top$) and ten-band ($bottom$) cases. The catastrophic outlier identification method discussed in \S \ref{gen} was employed for these determinations with the $\Delta{z_{peak,min}}=0.6$ and $p_{f,min} = 0.90$ criteria and the flagged galaxies are shown by red crosses. These determinations were performed with a training set consisting of 1200 galaxies chosen at random and an evaluation testing set consisting of the other 2504 galaxies. A bin size of 0.1 was used. Outliers in a determination are defined by equation \ref{erroreq}, shown as those points lying outside of the two diagonal lines. The density of points within the lines is quite high --- only 2.6\% of points lie outside of the lines as outliers for the ten-band case (BOTTOM) before flagging and 6.7\% for the five-band case (TOP).}
\label{f1_removed_red_xs}
\end{figure}

In order to form an additional test set with a significantly larger number of real galaxies, we also utilize galaxies from the COSMOS2015 photometric catalog that contain particularly reliable, previously estimated photometric redshifts derived from a large number of photometric bands.  As the COSMOS2015 catalog provides photometry for some galaxies in up to 31 optical, infrared, and UV bands, those galaxies with (i) magnitude values for at least 30 bands of photometry, and (ii) for which the stated $\chi^2$ for the redshift estimate is $<1$, and (iii) for which the stated photo-z value from the minimum $\chi^2$ estimate is less than 0.1 redshift away from the stated photo-z value from the peak of the PDF, can be considered to have highly reliable previous redshift estimates.  Applying these criteria result in a data set of 58622 galaxies spanning an $i$ band magnitude range from 27.17 to 19.00 with a median of 24.08.  For shorthand purposes we will refer to this set as the ``COSMOS-reliable-$z$'' test data set.  The $N(z)$ distribution for this data set is also shown in Figure \ref{p0}.

\begin{figure*}[!htb]
\includegraphics[width=.8\textwidth]{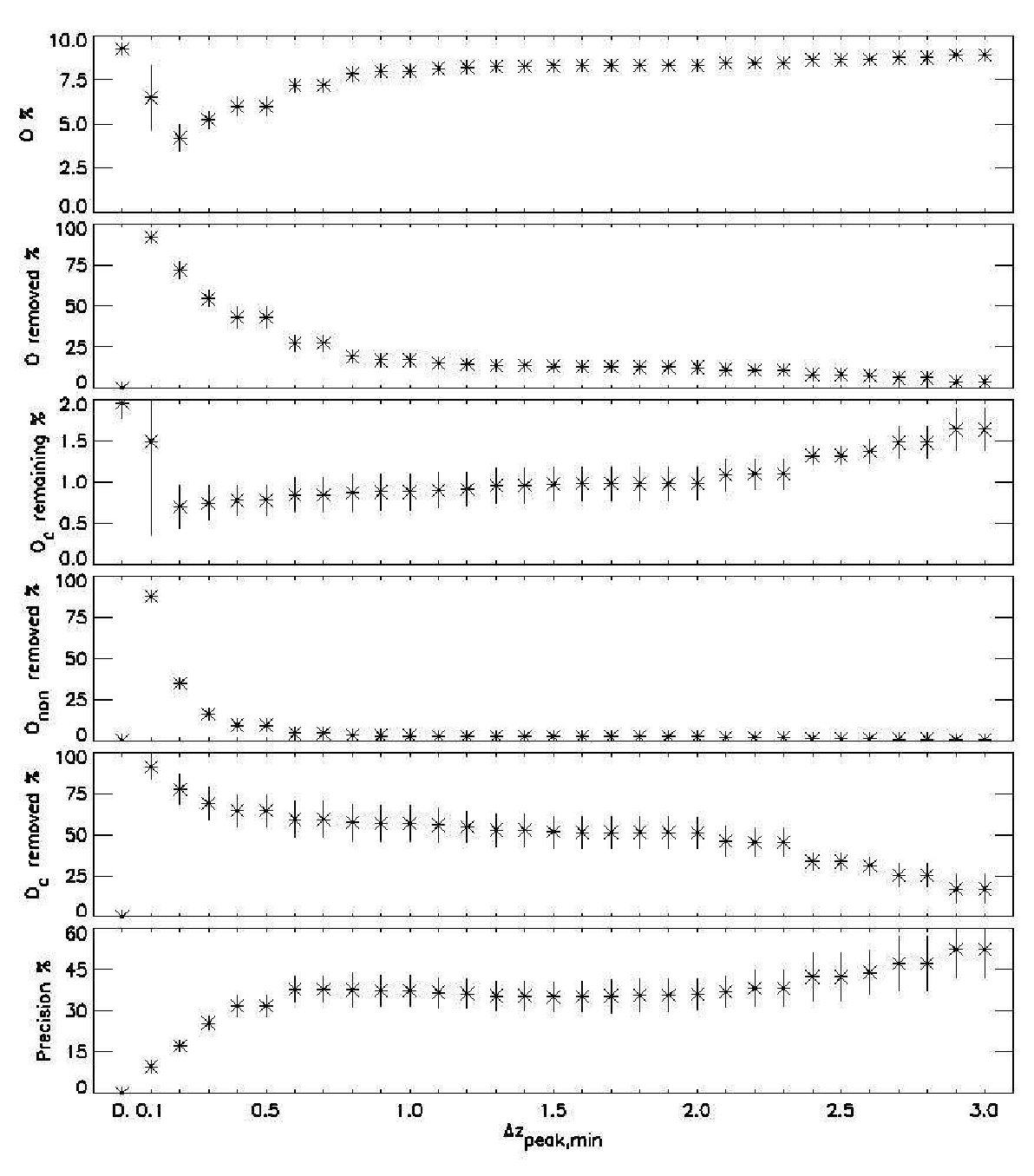}
\caption{Visualization of photo-z performance metrics from determinations performed by SPIDERz on the COSMOSx3D-HST data set discussed in \S \ref{data} for the five photometric band case using a range of $\Delta{z_{peak,min}}$ values and fixed $p_{f,min}$ = 0.90, considering that all flagged galaxies would be removed from an analysis that relied on accurate photo-zs. We also include the performance for the default case of no flagging on the left-most portion of the x-axis labeled ``D.''. The determinations were performed with a bin size of 0.1, and a training set consisting of 1200 galaxies chosen at random and an evaluation testing set consisting of the other 2504 galaxies, with results averaged over six determinations. The performance metrics shown include the percentage of outliers (TOP), followed by the percentage of outliers removed (2nd from TOP), followed by the percentage of catastrophic outliers remaining (3rd from TOP), followed by the percentage of non-outliers removed (3rd from BOTTOM), followed by the percentage of catastrophic outliers removed (2nd from BOTTOM), and finally the percentage of removed galaxies that are outliers (BOTTOM). The variance in performance across the six randomized realizations is indicated.}
\label{6stack}
\end{figure*}

\begin{figure*}[!htb]
\includegraphics[width=.8\textwidth]{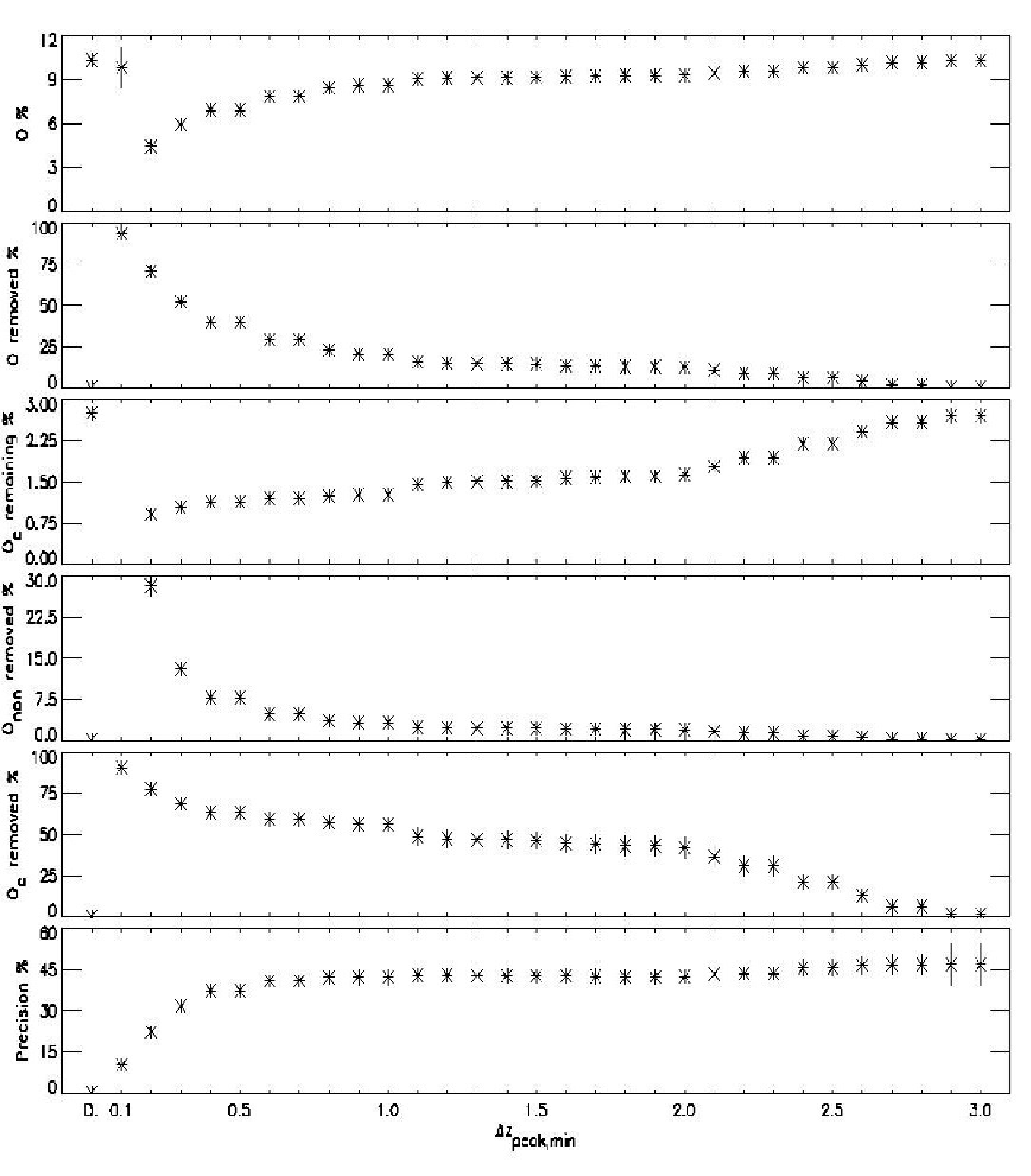}
\caption{Same as Figure \ref{6stack} but for the COSMOS-reliable-$z$ data set discussed in \S \ref{data}, and with fixed $p_{f,min}$ = 0.90. The variance in performance across the six randomized realizations is indicated.  Results from this data set are quite similar to those from the COSMOSx3D-HST data set shown in Figure \ref{6stack} but with smaller error bars, as would be expected from a much larger data set. }
\label{6stackz}
\end{figure*}

Although the COSMOS2015 catalog provides photometry in a potentially large number of optical, infrared, and UV bands, we choose to restrict our test analyses to the $u$, $B$, $V$, $r$, $i$, $z+$, $Y$, $H$, $J$, and $Ks$ bands, and a subset of five of these bands, because with data sets approaching 30 bands of photometry, the distinction between photo-z estimation and spectroscopic redshift determination is somewhat muddled, and in any case this does not represent a realistic photometric situation for upcoming large surveys such as LSST, even for subsets which would have infrared survey overlap.  In the following sections we refer to test data consisting of only five optical bands ($u$, $B$, $r$, $i$, $z+$) as the `five-band case', which could resemble the default situation for obtaining photometric redshifts from a very large optical survey, and similarly refer to test data comprised of all ten aforementioned bands as the `ten-band case', which could resemble the situation for obtaining photometric results from a large optical survey that overlaps a large near-infrared survey.  For these bands we use aperture magnitudes measured in a 3'' aperture.  The depths of the photometry for the bands are given in Table 1 of \cite{Laigle15}.  We have not utilized galaxies with missing photometry values in these bands --- for the COSMOSx3D-HST test set the number of galaxies where this is the case is negligible, while for the COSMOS-reliable-$z$ test set applying this filter has almost no effect since this data set by definition contains 30 reliable bands of photometry.  

Unless otherwise noted, all determinations are performed with randomly selected training and testing set populations of 1200 and 2504 galaxies respectively for the COSMOSx3D-HST data set, and 5000 and 53622 galaxies for the COSMOS-reliable-$z$ data set. Increasing the training population size beyond 1200 for the COSMOSx3D-HST data set produced only marginal improvements in photo-z accuracy.  For the COSMOS-reliable-$z$ data set we chose to maintain a training set to evaluation set size ratio of below 1:10 in order to more closely approximate the photo-z conditions of future large-scale survey analyses than would be achieved with doing analyses with larger ratios.

We note that the galaxies in these data sets span the largest redshift range of publicly available real galaxy photo-z test data with photometry down to these magnitudes of which we are aware. We also note that a significant limitation is posed on the performance accuracy of SPIDERz due to inadequate parameter overlap between training and evaluation galaxies in sparsely populated redshift regions, which, among other restrictive influences, imposes a lower limit on the redshift bin size that can be effectively used.

\subsection{Results for various parameter choices}\label{COSMOSres}

Figure \ref{f1_removed_red_xs} displays the estimated SPIDERz photo-z versus actual redshift for an example of typical determinations with the five-band and ten-band cases for the COSMOSx3D-HST data setdiscussed in \S \ref{data}. The EPDF outlier identification method discussed in \S \ref{gen} was then employed for these determinations with particular flagging parameters $\Delta{z_{peak,min}}=0.6$ and $p_{f,min}=0.95$. Red data points indicate flagged potential catastrophic outlier estimates in these cases.  Estimates with the ten-band case are of course significantly better than with the five-band case.

\begin{figure*}[!htb]
\includegraphics[width=0.5\textwidth,height=.4\textwidth]{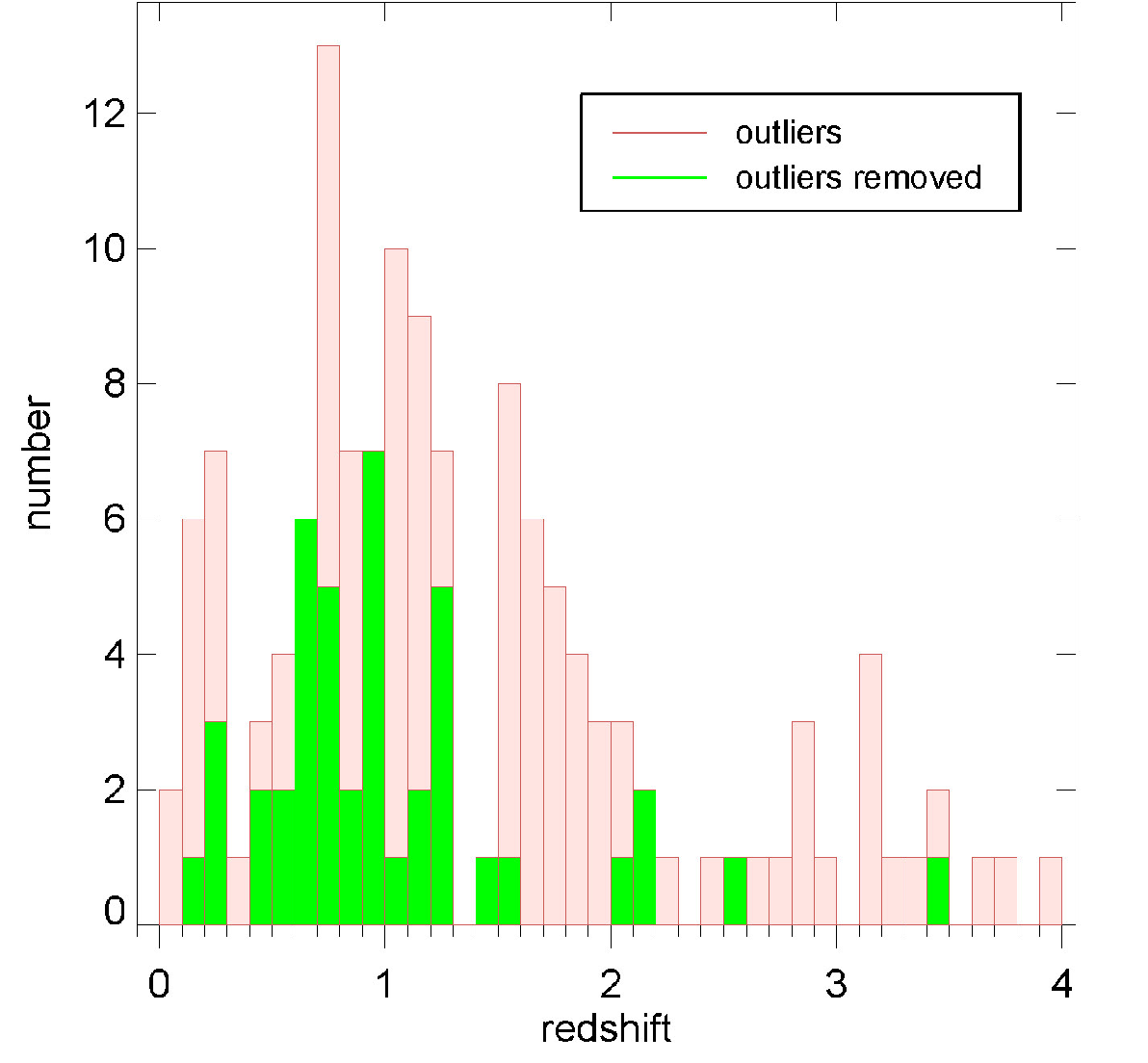}
\label{p2}
\includegraphics[width=0.5\textwidth,height=.4\textwidth]{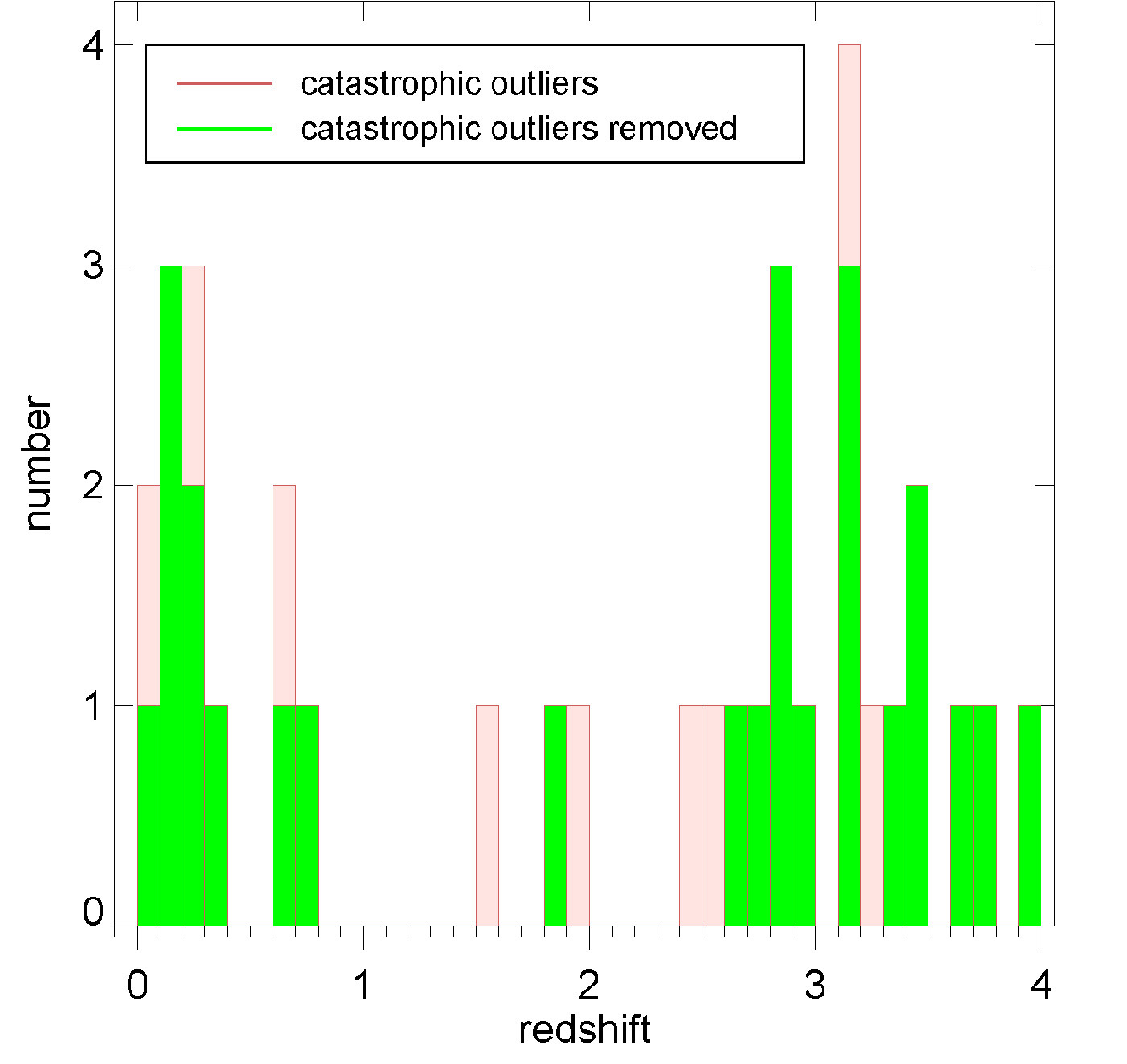}
\label{p4}
\caption{Redshift histogram of the number of outliers (left) and catastrophic outliers (right), both as defined in equations \ref{erroreq} and \ref{erroreqCO} respectively, present in one particular typical determination with the five photometric band case for the COSMOSx3DHST test data set compared to the numbers flagged through the use of the EPDF flagging method with flagging parameter values $p_{f,min} = 0.90$ and $\Delta{z_{peak,min}} = 0.6$.}
\label{p3}
\end{figure*}

To examine the influence of our proposed method for flagging potential catastrophic outliers in photo-z determinations, we performed an extensive analysis with test determinations on the five-band and ten-band cases for both the COSMOSx3D-HST and COSMOS-reliable-$z$ data sets using a range of $\Delta{z_{peak,min}}$ and $p_{f,min}$ values, redshift bin sizes, and training population sizes.

Perhaps surprisingly, we determine that appropriate values of $p_{f,min}$ are quite high, with any values below $p_{f,min} = 0.9$ resulting in an unacceptably large number of spurious flaggings. We find variations in the designated value for $\Delta{z_{peak,min}}$ greatly influence the performance of the outlier identification method, as measured by the relative numbers of correct outlier identifications versus spurious removal of non-outliers, however variations in $p_{f,min}$ produced marginal difference in the range $0.90 \leq p_{f,min} \leq 0.98$. 

We also find that discrete photo-z accuracy is generally highest on this test data when using redshift bin sizes between 0.1 and 0.05; the use of larger bin sizes significantly reduced photo-z precision across all $z$ values and particularly at lower $z$s, as expected, while the use of bin sizes less than 0.05 produced a significant number of unoccupied bins at higher redshifts and deteriorated parameter overlap between training and evaluation sets.

Figures \ref{6stack} and \ref{p3} and Tables \ref{tab2} and \ref{tab3} show various performance metrics from determinations with SPIDERz using the EPDF outlier identification method on COSMOSx3D-HST test data. Table \ref{tab2} highlights the percentage of outliers, percentage of outliers removed, percentage of removed galaxies that are outliers, percentage of non-outliers removed, percentage of catastrophic outliers removed, and finally the percentage of catastrophic outliers remaining for determinations on five-band and ten-band cases for this data set, with a range of values for $\Delta{z_{peak,min}}$ and a fixed $p_{f,min}$ of 0.90, while figure \ref{6stack} provides a visual compendium of some of those quantities for the five-band case. Table \ref{tab3} shows various metrics for several combinations of $\Delta{z_{peak,min}}$ and $p_{f,min}$ values.  Figure \ref{p3} shows a redshift histogram of the reduction in the number of catastrophic outliers and outliers present in a typical determination with the five-band case with one particular parameter value choice.  Figure \ref{6stackz} shows performance metrics from determinations with SPIDERz using the EPDF outlier identification method on the COSMOS-reliable-z test data set with a fixed $p_{f,min}$ of 0.90.  Comparing Figures \ref{6stack} and \ref{6stackz} it is clear that results from the two test data sets are quite similar but with significantly smaller error bars in the COSMOS-reliable-$z$ case, as would be expected from a significantly larger data set.

We see that certain choices for $\Delta{z_{peak,min}}$ and $p_{f,min}$ result in successfully flagging a high percentage ($>$ 50\%) of the catastrophic outliers while flagging a small percentage (2-4\%) of the non-outlier galaxies. On the other hand, low values of $\Delta{z_{peak,min}}$ result in the flagging of a large percentage of the non-outlier galaxies.

It is also of interest to explore whether this method flags an excessive fraction of galaxies at higher redshifts and/or higher magnitudes.  In Figure \ref{n_i} we show the percentage of non-outliers flagged in bins of 0.1 in redshift (left panel) and in sextiles of $i$-band magnitude (right panel) for the COSMOS-reliable-$z$ test data set with flagging parameter values $p_{f,min} = 0.95$ and $\Delta{z_{peak,min}} = 0.6$.  It is seen that less than 15\% of non-outliers are flagged in the highest magnitude (dimmest flux) sextile, but in a few of the least populated redshift bins in the sample roughly half of non-outliers are flagged.  This suggests that steps could be taken to mitigate this effect within certain low population redshift bins, as discussed in \S \ref{disc}.

\section{Discussion}\label{disc}

In this work we have considered the utilization of SPIDERz's effective redshift probability distributions for flagging likely catastrophic outlier photo-z predictions --- gross mis-estimations defined by $|z_{phot} - z_{spec}| > 1$ --- by considering galaxies with multiple or ill-defined peaks in photo-z probability separated by redshift. We introduced a formalism with two threshold criteria: the minimum redshift separation of multiple peaks ($\Delta{z_{peak,min}}$) and the minimum probability ratio of secondary probability peaks to the highest probability peak ($p_{f,min}$), as discussed in \S \ref{gen}, to preemptively flag potential catastrophic outlier estimates. We implemented this method in SPIDERz photo-z determinations performed with real galaxy test data spanning a wide redshift range $0 < z < 4$ and utilizing limited photometric bands to estimate photometric redshift (see \S \ref{data}), testing a range of threshold values for $\Delta{z_{peak,min}}$ and $p_{f,min}$.

We found $\Delta{z_{peak,min}}$ to have the greatest influence on the fraction of catastrophic outliers which were flagged, while $p_{f,min}$ was sub-dominant in this regard but most strongly correlated with flagging precision, with low values of $p_{f,min}$ leading to a higher number of non-outliers flagged. Optimal values for $\Delta{z_{peak,min}}$ and $p_{f,min}$ for any given application would result from striking an acceptable balance between more thoroughly flagging catastrophic outlier galaxies and reducing the number of spuriously flagged non-outlier galaxies. 

\begin{figure}
\resizebox{\hsize}{!}{\includegraphics{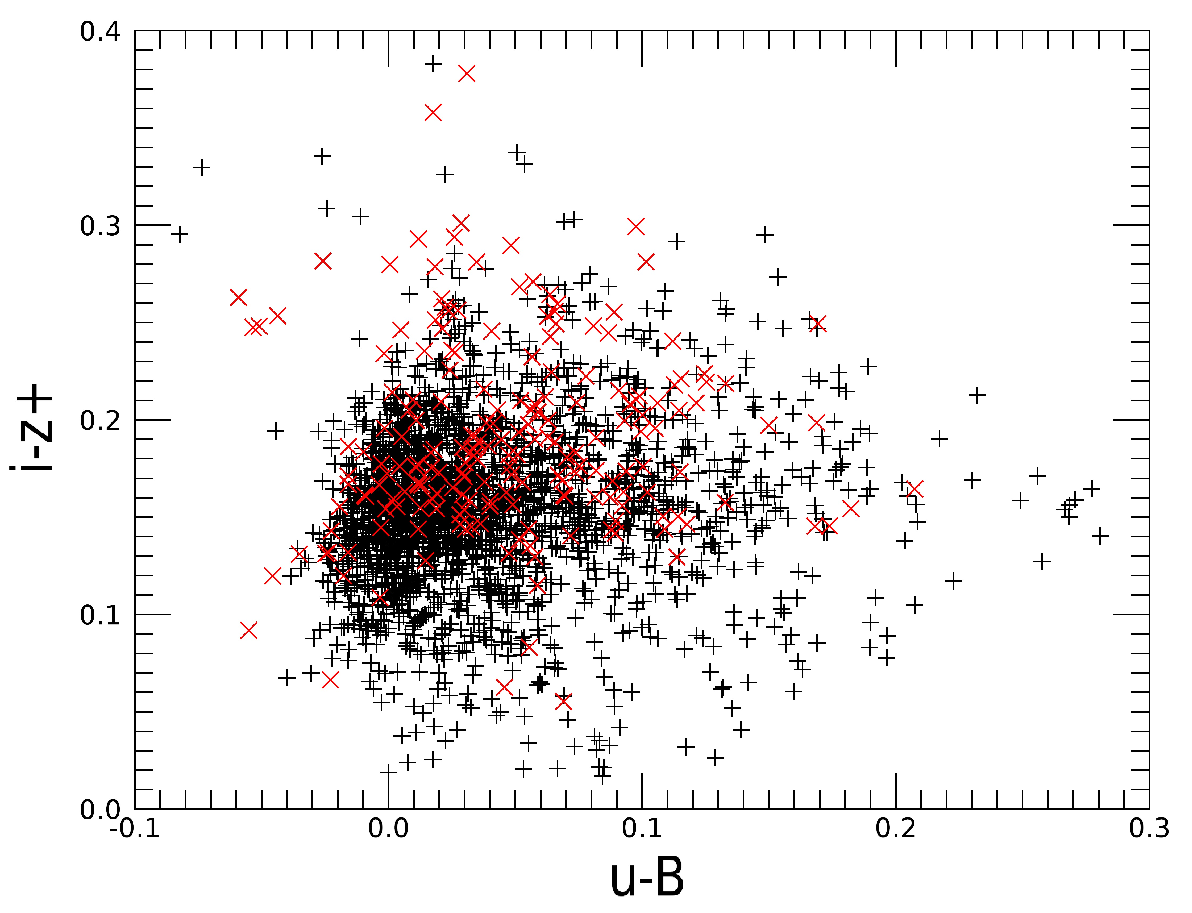}}
\caption{Galaxies of the COSMOSx3D-HST test data set plotted in a two-color space, with those galaxies flagged under the optimized $\Delta{z_{peak,min}} = 0.6$ and $p_{f,min} = 0.90$ criteria indicated with red crosses.  In this analysis we find that flagged galaxies do not occupy systematically different regions of color space than non-flagged galaxies, indicating that flagging does not favor one particular galaxy type or types in this case.}
\label{colorplot}
\end{figure}

We present results for a variety of choices of $\Delta{z_{peak,min}}$ where this trade-off can be seen, particularly in Figure \ref{6stack} and Tables 1 and 2. There are a range of values for $\Delta{z_{peak,min}}$ where the percentage of catastrophic outliers flagged is quite high and the percentage of non-outliers flagged is relatively low. For all parameter choices, more non-outliers are flagged than outliers, but this is likely inevitable considering that in the default case the vast majority of galaxies are non-outliers ($>$ 90\% of the galaxies in the five-band case and $>$ 95\% in the ten-band case).

\begin{figure*}[!htb]
\includegraphics[width=0.5\textwidth]{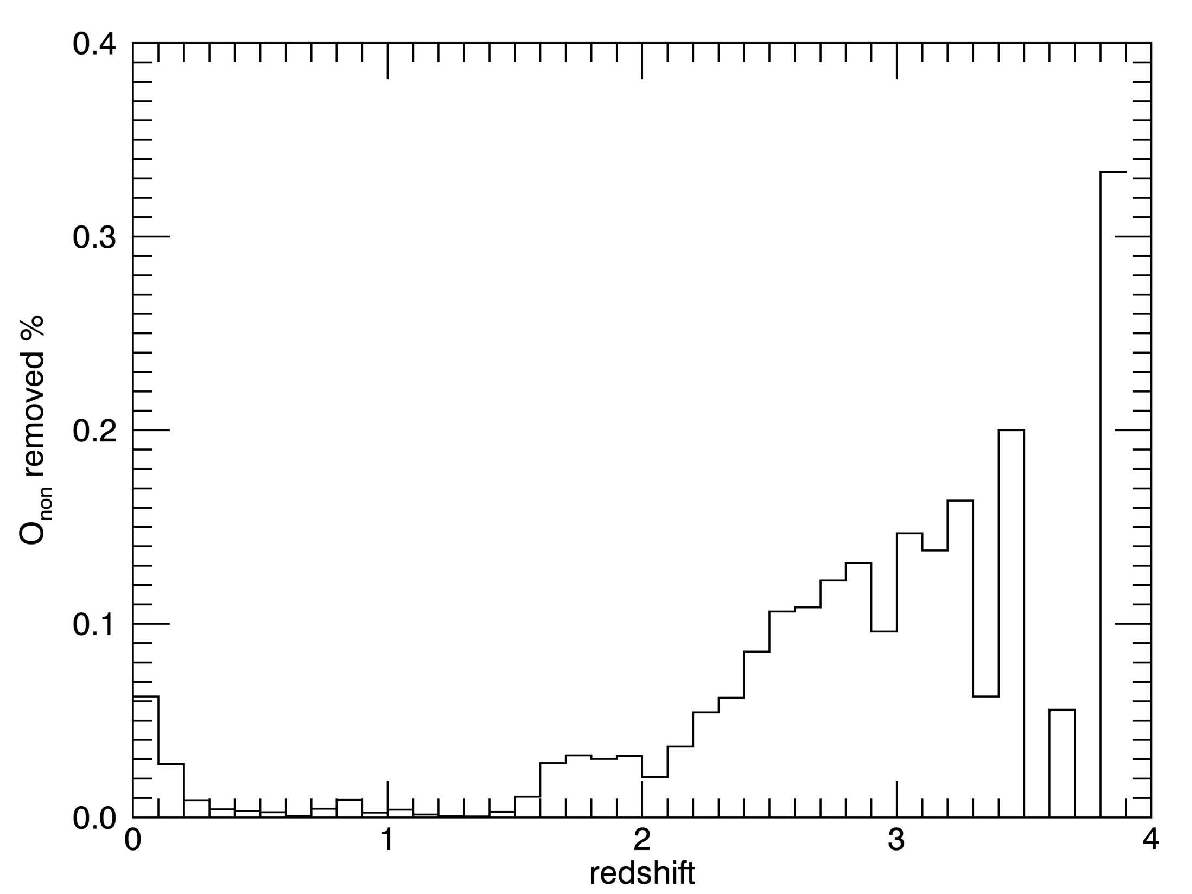}
\label{z_rems}
\includegraphics[width=0.48\textwidth]{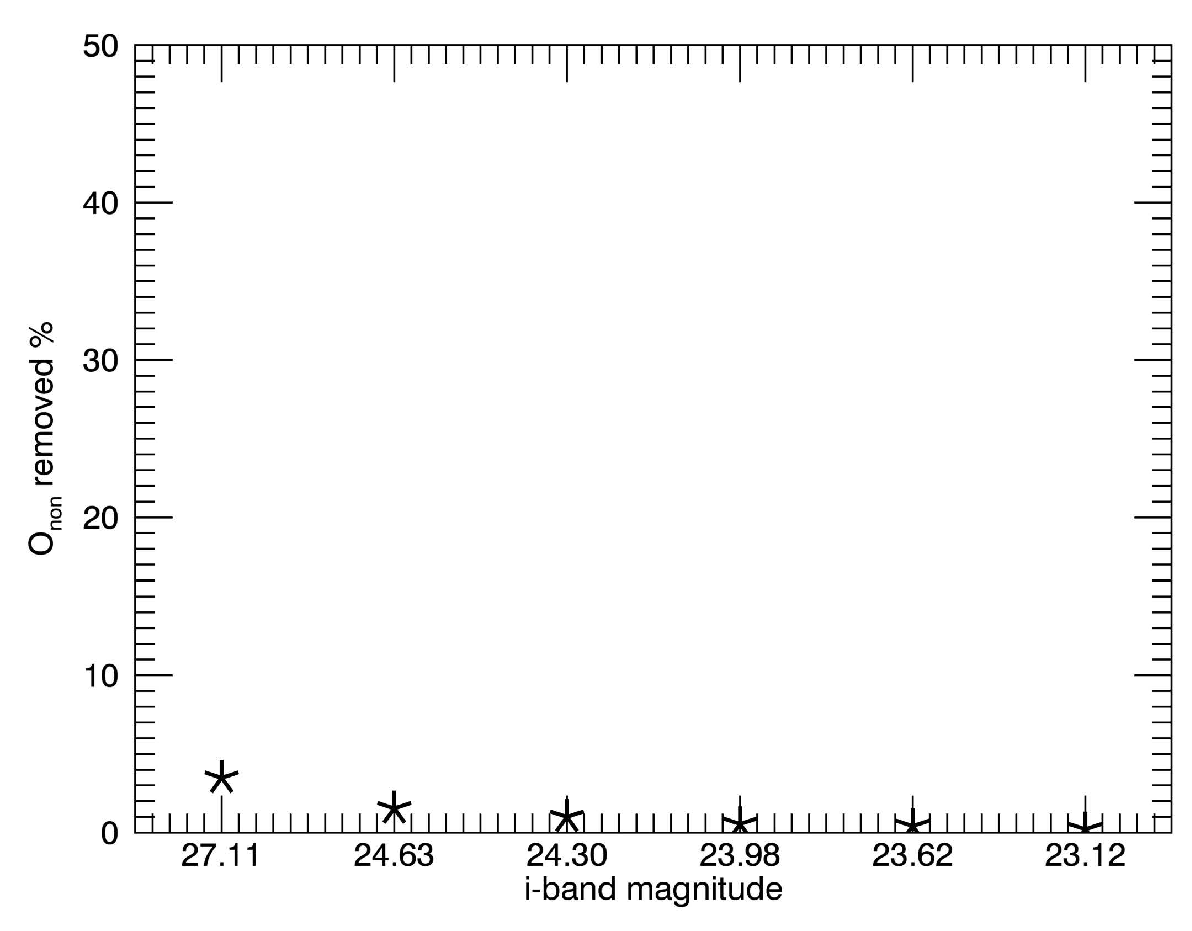}
\caption{The percentage of non-outliers flagged through the use of the EPDF flagging method in bins of 0.1 in redshift (left) and in equally populated sextiles of $i$-band magnitude from highest to lowest magnitude indexed with the median magnitude (right) for the COSMOS-reliable-$z$ test data set with flagging parameter values $p_{f,min} = 0.90$ and $\Delta{z_{peak,min}} = 0.6$.  The results here are for one particular representative determination.  The standard deviation from averaging over multiple determinations would be smaller than the plotting symbols in the $i$-band case and small in the redshift case.  The redshift bins where a large fraction of non-outliers are flagged are those which are least populated in the sample generally.}
\label{n_i}
\end{figure*}

We have seen that with proper choices for $\Delta{z_{peak,min}}$ and $p_{f,min}$, EPDFs can be utilized to flag potential catastrophic outlier photo-z predictions with a high degree of overall effectiveness in determinations performed on a data set which spans a wide redshift range and contains realistic photometry in a limited number of wavebands. As discussed in \S \ref{gen}, in a future large-scale survey utilizing photometric redshifts, the simplest use of such flagging information would be to simply remove the flagged galaxies from science analyses in which catastrophic outlier redshift predictions are detrimental, such as weak-lensing cosmology. Another simple option for utilization of flagging information could include de-weighting of potential catastrophic outliers in cosmological probes.

If such flagged galaxies are simply removed from analysis, there is, necessarily, a trade-off between more complete removal of actual catastrophic outliers and spurious removal of non-outliers. In this work we present various options for the parameters $\Delta{z_{peak,min}}$ and $p_{f,min}$ (discussed in \S \ref{gen}) which lead to different points on this trade-off continuum. We show the various results for catastrophic outliers removed, spurious removals, and other metrics in Tables \ref{tab2} and \ref{tab3} and visualizations in Figures \ref{p3} and \ref{6stack}. It is seen that for a range of flagging parameter values a favorable ratio of total genuine catastrophic outlier flagging to spurious non-outlier flagging is obtained, for example flagging of significantly more than half of catastrophic outliers while spuriously flagging only 2-4\% of non-outliers.  With the need to obtain precise redshift estimates satisfying photo-z error constraints for probing cosmological parameters and the abundance of galaxies that will be observed in future large photometric surveys, it may be reasonable in many cases to accept a slightly larger (although still low) percentage of overall spurious removals in exchange for maximizing the number of removed catastrophic outlier photo-z estimates.

One could ask whether such removal or de-weighting could preferentially remove galaxies of a certain type or types.  While this should have only second-order effects on cosmological probes (assuming a scenario where a purely empirical photo-z estimation technique is used), it may be of concern for astrophysical studies.  In this analysis we find that flagged galaxies do not differ systematically in colors from non-flagged galaxies.  Figure \ref{colorplot} shows an example photo-z determination with the COSMOSx3D-HST data set in a two-color space which used the same optimized flagging parameters for the determination shown in Figure \ref{f1_removed_red_xs}.  The extent to which this generalizes to any data set is not known at present.  However with any given data set it could be evaluated by testing photo-z evaluation and flagging on a sub-sample of galaxies of known redshift, such as a part of the spectroscopic training set itself.

It is important to note, however, that as seen in Figure \ref{n_i} a significant fraction of non-outliers are flagged in a few of the more sparsely populated redshift bins, including some of those at higher redshifts.  This points toward a possible strategy beyond simple removal of flagged galaxies in these particular redshift bins in order to not lose for cosmological analyses such a large fraction of high redshift galaxies in a data set.  We will explore possible weighting strategies for this in a future work.  We do also note two crucial caveats regarding this: (1) that in this work, as mentioned in \S \ref{intro}, in order to approximate the photo-z conditions applying to future large-scale surveys, we utilize much larger evaluation sets than training sets in this study.  Thus it is likely that by adopting a larger training to evaluation set size ratio than here, as has been done in many other photo-z studies in the literature, one could reduce the percentage of spuriously flagged non-outliers in the sparsely populated redshift bins given a similarly sized test data set.  Also, (2) it is likely the case that, for a given training to evaluation set size ratio and $N(z)$ distribution, there will be a lower percentage of spuriously flagged non-outliers in relatively sparsely populated redshift bins given a larger overall test data set.  However even with a very large training set high redshift bins will contain a higher proportion of potential catastrophic outliers and therefore spurious removals due to the degeneracy between Balmer and Lyman breaks in galaxy spectra.

While this analysis focused on utilization of EPDFs provided by SPIDERz, there is no reason that it should not in principle be generalizable with analogous parameters to any photo-z estimation method which provides redshift probability distribution information for each galaxy.  We note though that the specific optimal flagging parameter values might be quite different for a non-empirical photo-z estimation method. While the parameters we used in this work to flag EPDF features, $\Delta{z_{peak,min}}$ and $p_{f,min}$, were effective in distinguishing likely catastrophic outliers, the optimal values of these parameters for a given purpose are data set dependent to some extent.  Also other photo-z estimation codes and probability determination methods may or may not necessitate alternate parameter values and/or definitions to those employed in this work.  We also note that in general results in empirical photo-z estimation methods often depend on the degree of representativeness of the training set relative to the evaluation set.

\begin{table*}[!htb]
\caption{Results for analyses performed with SPIDERz on the five and ten photometric band test data sets derived from the COSMOSx3D-HST data discussed in \S \ref{data}. Determinations feature 1200 galaxies used for training and the remaining 2504 galaxies used for evaluation. Six determinations were performed for every case, each with randomized training and evaluation testing sets, and results averaged. Results are shown for the default cases with no flagging, and also with implementation of the EPDF outlier flagging method discussed in \S \ref{gen} using a range of $\Delta{z_{peak,min}}$ values and fixed $p_{f}$ = 0.90, assuming that all flagged galaxies would be removed from a data set that relied on accurate photo-zs, to illustrate the percentage reduction in outlier and catastrophic outlier galaxies achieved at the cost of incorrectly removing a percentage of non-outlier galaxies. Here we use the shorthand $O$ and $O_c$ for outliers and catastrophic outliers, respectively, which are defined by equations \ref{erroreq} and \ref{erroreqCO}, and $O_{non}$ for non-outliers. The `Precision' refers to the percentage of flagged galaxies which are outliers. The RMS and reduced RMS errors are also included for each case and defined by equation \ref{RMSeq} as discussed in \S \ref{intro}.}
\label{tab2}
\begin{center}
\resizebox{\textwidth}{!}{
\begin{tabular}{lrrrrrrrr}
\hline
\hline
$\Delta{z_{peak,min}}$&$O$ \% & $O^{{removed}}$ \% & $O_{c}$ \% & $O^{{removed}}_c$ \% & $O^{{removed}}_{{non}}$ \% & Precision \% & $\sigma_{RMS}$ & $\sigma_{R-RMS}$\\
\hline
Five photometric bands \\ (0.1 bin size)\\
\hline
Default & 11.6 & - & 3.17 & - & - & - & 0.222 & 0.052\\
\hline
0.2  & 4.81 & 58.5 & 0.949 & 81.2 & 32.2 & 23.2 & 0.118 & 0.035  \\
0.3  & 6.12 & 47.2 & 1.04 & 74.5 & 16.9 & 31.4 & 0.129  &  0.042 \\
0.4  & 7.19 & 38.0 & 1.15 & 69.5 &  11.5 &  35.3 & 0.141 &  0.045 \\
0.5  & 7.19 & 38.0 & 1.15 & 69.5 & 11.5 & 35.3 & 0.141 &  0.045 \\
0.6  & 8.46 & 27.1 & 1.31 & 63.0 & 6.95 &  39.4& 0.157 &  0.048 \\
0.7  & 8.46 & 27.1 & 1.31 & 63.0 & 6.95 &  39.4 &  0.157 & 0.048 \\
0.8  & 9.29 & 19.9 & 1.37 & 59.9 &  4.75& 41.5 & 0.161 & 0.049  \\
0.9  & 9.57 & 17.5 & 1.43 & 57.8 & 4.18 & 41.6  &  0.162 & 0.049 \\
1.0  & 9.57 & 17.5 & 1.43 & 57.8 &  4.18 &  41.6 &  0.162 & 0.049 \\
\hline
Ten photometric bands\\ (0.05 bin size) \\
\hline
Default & 4.17 & - & 1.08 & - & - & - & 0.144 & 0.047\\
\hline
0.2 & 0.938 & 77.5 & 0.146 & 86.5 & 53.1 & 6.89 & 0.064 & 0.025\\
0.3 & 1.49 & 64.3 & 0.129 & 88.1 & 20.5 & 13.4 & 0.069 & 0.037\\
0.4 & 2.02 & 51.6 & 0.180 & 83.3 & 9.59 & 20.8 & 0.078 & 0.042\\
0.5 & 2.21 & 47.0 & 0.198 & 81.7 & 7.48 & 23.4 & 0.079 & 0.043\\
0.6 & 2.66 & 36.2 & 0.218 & 79.8 & 3.96 & 30.8 & 0.085 & 0.045\\
0.7 & 2.76 & 33.8 & 0.237 & 78.1 & 3.44 & 32.3 & 0.087 & 0.045\\
0.8 & 2.95 & 29.3 & 0.257 & 76.2 & 2.92 & 33.0 & 0.088 & 0.046\\
0.9 & 3.07 & 26.4 & 0.277 & 74.4 & 2.74 & 32.3 & 0.090 & 0.046\\
1.0 & 3.12 & 25.2 & 0.290 & 73.1 & 2.66 & 32.1 & 0.090 & 0.046\\
\hline
\end{tabular}
}
\end{center}
\end{table*}

\begin{table*}[!htb]
\caption{Improvements in RMS and R-RMS (defined by equation \ref{RMSeq}), and the percentage of catastrophic outliers ($O_c$, defined in equation \ref{erroreqCO}) after flagging potential catastrophic outlier EPDFs in SPIDERz determinations on COSMOSx3D-HST test data for the five photometric band case for a range of $\Delta{z_{peak,min}}$ and $p_{f,min}$ values, with a redshift bin size of 0.1, assuming removal of flagged galaxies, where $\Delta\%$ indicates a comparison to default values. We section results according to $\Delta{z_{peak,min}}$ values and recomputed default determinations in each section. Six determinations were performed for every case, each with randomized training and evaluation testing sets consisting of 1200 and 2504 galaxies respectively, and results averaged. We also show the percentage of non-outliers ($O_{non}$) flagged.}
\label{tab3}
\begin{center}
\tiny
\resizebox{\textwidth}{!}{
\begin{tabular}{lrrrrrrrr}
\hline
\hline
 $p_{f,min}$ & $\Delta{z_{peak,min}}$ & $O_{c}$ \% & $\Delta\%$& $\sigma_{RMS}$ & $\Delta\%$ & $\sigma_{R-RMS}$ & $\Delta\%$& $O^{{removed}}_{{non}}$ \% \\
\hline
Default	&	-	&	2.90	&	-	&	0.225	&	-	&	0.052	&	-	&	-	\\
0.98	&	0.2	&	2.63	&	-0.094	&	0.208	&	-0.077	&	0.050	&	-0.034	&	2.48	\\
0.98	&	0.3	&	2.64	&	-0.090	&	0.209	&	-0.071	&	0.051	&	-0.021	&	1.50	\\
0.98	&	0.4	&	2.64	&	-0.090	&	0.209	&	-0.071	&	0.051	&	-0.021	&	1.50	\\
0.98	&	0.5	&	2.65	&	-0.086	&	0.210	&	-0.066	&	0.051	&	-0.014	&	0.937	\\
0.98	&	0.6	&	2.65	&	-0.086	&	0.210	&	-0.066	&	0.051	&	-0.008	&	0.937	\\
0.98	&	0.7	&	2.65	&	-0.084	&	0.211	&	-0.064	&	0.051	&	-0.008	&	0.545	\\
0.98	&	0.8	&	2.66	&	-0.083	&	0.211	&	-0.062	&	0.051	&	-0.006	&	0.364	\\
\hline
Default	&	-	&	3.04	&	-	&	0.224	&	-	&	0.051	&	-	&	-	\\
0.95	&	0.2	&	1.98	&	-0.348	&	0.176	&	-0.216	&	0.045	&	-0.115	&	9.21	\\
0.95	&	0.3	&	2.02	&	-0.335	&	0.180	&	-0.196	&	0.048	&	-0.066	&	5.18	\\
0.95	&	0.4	&	2.05	&	-0.326	&	0.183	&	-0.182	&	0.049	&	-0.043	&	3.20	\\
0.95	&	0.5	&	2.05	&	-0.326	&	0.183	&	-0.182	&	0.049	&	-0.043	&	3.20	\\
0.95	&	0.6	&	2.07	&	-0.318	&	0.185	&	-0.172	&	0.050	&	-0.028	&	2.00	\\
0.95	&	0.7	&	2.07	&	-0.318	&	0.185	&	-0.172	&	0.050	&	-0.028	&	2.00	\\
0.95	&	0.8	&	2.09	&	-0.311	&	0.186	&	-0.168	&	0.050	&	-0.022	&	1.60	\\
\hline
Default	&	-	&	3.17	&	-	&	0.222	&	-	&	0.052	&	-	&	-	\\
0.9	&	0.2	&	0.949	&	-0.700	&	0.118	&	-0.471	&	0.035	&	-0.334	&	32.2	\\
0.9	&	0.3	&	1.04	&	-0.671	&	0.129	&	-0.419	&	0.042	&	-0.192	&	16.7	\\
0.9	&	0.4	&	1.16	&	-0.635	&	0.141	&	-0.365	&	0.045	&	-0.129	&	11.5	\\
0.9	&	0.5	&	1.16	&	-0.635	&	0.141	&	-0.365	&	0.045	&	-0.129	&	11.5	\\
0.9	&	0.6	&	1.31	&	-0.587	&	0.157	&	-0.293	&	0.048	&	-0.077	&	6.95	\\
0.9	&	0.7	&	1.31	&	-0.587	&	0.157	&	-0.293	&	0.048	&	-0.077	&	6.95	\\
0.9	&	0.8	&	1.37	&	-0.567	&	0.161	&	-0.276	&	0.049	&	-0.053	&	4.75	\\
\hline
Default	&	-	&	3.19	&	-	&	0.227	&	-	&	0.052	&	-	&	-	\\
0.8	&	0.2	&	1.36	&	-0.576	&	0.099	&	-0.565	&	0.006	&	-0.880	&	97.6	\\
0.8	&	0.3	&	0.454	&	-0.858	&	0.085	&	-0.624	&	0.020	&	-0.618	&	76.1	\\
0.8	&	0.4	&	0.463	&	-0.855	&	0.101	&	-0.557	&	0.030	&	-0.417	&	52.9	\\
0.8	&	0.5	&	0.463	&	-0.855	&	0.101	&	-0.557	&	0.030	&	-0.417	&	52.9	\\
0.8	&	0.6	&	0.562	&	-0.824	&	0.119	&	-0.478	&	0.040	&	-0.230	&	30.6	\\
0.8	&	0.7	&	0.562	&	-0.824	&	0.119	&	-0.478	&	0.040	&	-0.230	&	30.6	\\
0.8	&	0.8	&	0.631	&	-0.802	&	0.126	&	-0.446	&	0.044	&	-0.146	&	19.2	\\
\hline
Default	&	-	&	2.94	&	-	&	0.223	&	-	&	0.052	&	-	&	-	\\
0.7	&	0.2	&	-	&	-	&	-	&	-	&	-	&	-	&	100	\\
0.7	&	0.3	&	1.69	&	-0.425	&	0.058	&	-0.739	&	0.004	&	-0.915	&	99.9	\\
0.7	&	0.4	&	0.171	&	-0.942	&	0.061	&	-0.725	&	0.010	&	-0.799	&	98.7	\\
0.7	&	0.5	&	0.171	&	-0.942	&	0.061	&	-0.725	&	0.010	&	-0.799	&	98.7	\\
0.7	&	0.6	&	0.231	&	-0.922	&	0.088	&	-0.604	&	0.024	&	-0.536	&	87.4	\\
0.7	&	0.7	&	0.231	&	-0.922	&	0.088	&	-0.604	&	0.024	&	-0.536	&	87.4	\\
0.7	&	0.8	&	0.256	&	-0.913	&	0.100	&	-0.551	&	0.035	&	-0.320	&	66.2	\\
\hline
\end{tabular}
}
\end{center}
\end{table*}

\end{document}